\documentclass[aps,prd,twocolumn,nofootinbib,superscriptaddress]{revtex4-1}
\usepackage{graphicx}  
\usepackage{dcolumn}   
\usepackage{bm}        
\usepackage{amssymb,amsfonts,amsmath,physics}   
\usepackage{cancel}
\usepackage{mathtools}
\usepackage{slashed}
\usepackage[dvipsnames]{xcolor}

\usepackage{tikz-feynman}
\tikzfeynmanset{compat=1.1.0}
\usetikzlibrary{arrows}
\tikzset{	
	vertex/.style={circle,draw, minimum size=1.5em},	
	edge/.style={->,> = latex'}	
}
\usepackage[bookmarks, breaklinks, colorlinks,urlcolor=blue, citecolor=red, 
linkcolor=blue]{hyperref}
\usepackage[normalem]{ulem}

\hypersetup{
	colorlinks=true,
	linkcolor=Red,
	filecolor=magenta,
	urlcolor=Blue,
	citecolor=PineGreen}

\newcommand{\be}{\begin{eqnarray*}}
	\newcommand{\ee}{\end{eqnarray*}}

\newcommand{\bee}{\begin{eqnarray}}
	\newcommand{\eee}{\end{eqnarray}}
\newcommand{\beeq}{\begin{equation}}
	\newcommand{\eeq}{\end{equation}}

\usepackage{xspace}

\newcommand{\ba}{\begin{array}}
	\newcommand{\ea}{\end{array}}
\newcommand{\bd}{\begin{displaymath}}
	\newcommand{\ed}{\end{displaymath}}
\newcommand{\besub}{\begin{subequations}}
	\newcommand{\eesub}{\end{subequations}}
\newcommand{\bea}{\begin{eqnarray}}
	\newcommand{\eea}{\end{eqnarray}}

\newcommand{\sla}[1]{/\!\!\!#1}

\def\q2 {q^2}

\usepackage{tikz}
\usetikzlibrary{arrows,shapes}
\usetikzlibrary{trees}
\usetikzlibrary{matrix,arrows} 				
\usetikzlibrary{positioning}				
\usetikzlibrary{calc,through}				
\usetikzlibrary{decorations.pathreplacing}  
\usepackage{pgffor}							

\usetikzlibrary{decorations.pathmorphing}	
\usetikzlibrary{decorations.markings}
\tikzset{
	vector/.style={decorate, decoration={snake}, draw},
	provector/.style={decorate, decoration={snake,amplitude=2.5pt}, draw},
	antivector/.style={decorate, decoration={snake,amplitude=-2.5pt}, draw},
	fermion/.style={draw=black, postaction={decorate},
		decoration={markings,mark=at position .55 with {\arrow[draw=black]{>}}}},
	fermionbar/.style={draw=black, postaction={decorate},
		decoration={markings,mark=at position .55 with {\arrow[draw=black]{<}}}},
	fermionnoarrow/.style={draw=black},
	gluon/.style={decorate, draw=black,
		decoration={coil,amplitude=4pt, segment length=5pt}},
	scalar/.style={dashed,draw=black, postaction={decorate},
		decoration={markings,mark=at position .55 with {\arrow[draw=black]{>}}}},
	scalarbar/.style={dashed,draw=black, postaction={decorate},
		decoration={markings,mark=at position .55 with {\arrow[draw=black]{<}}}},
	scalarnoarrow/.style={dashed,draw=black},
	electron/.style={draw=black, postaction={decorate},
		decoration={markings,mark=at position .55 with {\arrow[draw=black]{>}}}},
	bigvector/.style={decorate, decoration={snake,amplitude=4pt}, draw},
}

\tikzstyle{block} = [draw, rectangle, 
minimum height=3em, minimum width=6em]

\begin{document}

\title{Spontaneous Leptogenesis with sub-GeV Axion Like Particles}

\author{Arghyajit Datta}
\email{arghyad053@gmail.com}
\affiliation{Center for Precision Neutrino Research, Chonnam National University, Gwangju 61186, Republic of Korea }
\affiliation{Department of Physics, Kyungpook National University, Daegu 41566, Republic of Korea}

\author{Soumen Kumar Manna}
\email{skmanna2021@gmail.com}
\affiliation{Department of Physics, Indian Institute of Technology Guwahati, Assam-781039, India}

\author{Arunansu Sil}
\email{asil@iitg.ac.in}
\affiliation{Department of Physics, Indian Institute of Technology Guwahati, Assam-781039, India}

\begin{abstract} 


A derivative coupling of an axion like particle (ALP) with a $B-L$ current may lead to the baryon asymmetry of the Universe via spontaneous leptogenesis provided a lepton number breaking interaction prevails in thermal equilibrium. Conventionally, such scenario works only for heavy ALPs and high reheating temperature due to the fact that the same lepton number breaking contribution is tied up with neutrino mass generation also. In this work, we propose inert Higgs doublet assisted lepton number violating operator to relieve such tension so as to generate lepton asymmetry (of freeze-in/out type) with a much lower reheating temperature that can accommodate light (sub-GeV) ALPs sensitive to current and future ALP searches. 


\end{abstract}
\maketitle

\section{Introduction}
The QCD axion, originally introduced by Peccei and Quinn for a dynamical resolution to the strong CP problem, carries resemblance to the Nambu Goldstone boson of a spontaneously broken global symmetry $U(1)_{\rm{PQ}}$. Its coupling 
with the CP {\it{violating}} topological gluon density $\phi G \tilde{G}/f_\phi$, suppressed by the scale of spontaneous symmetry breaking $f_\phi$, is of central importance in this context \cite{Peccei:1977hh,Peccei:1977ur,Weinberg:1977ma,Wilczek:1977pj,Kim:1979if,Shifman:1979if,Zhitnitsky:1980tq,Dine:1981rt}. It can also couple to other Standard Model (SM) gauge bosons via chiral anomaly. A more general class of models exists where the pseudo Nambu-Goldstone bosons associated to the spontaneous breaking of a global symmetry, other than $U(1)_{\rm{PQ}}$, are prevalent \cite{Svrcek:2006yi,Arvanitaki:2009fg}. Similar to the QCD axion, such axion-like particles (ALP) can interact 
with the SM gauge fields through dimension-5 operator: $\phi F\tilde{F}/f_\phi$ with $F$ being the SM gauge
field strength, while a shift symmetric derivative coupling of ALPs with the SM fermion current such as Baryon ($j^{\mu}_B$) or Lepton ($j^{\mu}_L$) current of the form $\partial_\mu \phi j^\mu_X/f_\phi$ may also be introduced at the effective level. 

Being feebly coupled to the SM fields and the presence of arbitrariness involved in its decay constant and mass in a wide possible range, the ALP phenomenology turns out to be quite rich and intriguing \cite{Preskill:1982cy,Abbott:1982af,Dine:1982ah,Arias:2012az,Marsh:2015xka,DiLuzio:2020wdo,Manna:2023zuq,OHare:2024nmr} in explaining some of the unresolved problems of particle physics and cosmology. For example, the above mentioned derivative coupling of  ALPs with $j^\mu_B$ is capable of explaining the baryon asymmetry of the Universe (BAU) via spontaneous baryogenesis \cite{Cohen:1987vi,Cohen:1988kt}. Considering the ALP field being homogeneous in space (happens to be the case provided the global $U(1)$ symmetry breaks spontaneously before inflation), a dynamic CPT violation 
results via $(\partial_0 \phi) j^0_B/f_\phi$ once the ALP attains a nonzero velocity $\dot{\theta}$, where $\theta= {\phi}/{f_\phi}$. 
Such a situation accompanied by a baryon number violating interaction in thermal equilibrium can be responsible for generation of BAU while disregarding the Sakharov's third condition \cite{Sakharov:1967dj}. 


A similar approach can also be exercised by replacing $j^\mu_B$ with the SM lepton current $j^\mu_L$ in the CPT 
violating source term, thereby realising spontaneous leptogenesis \cite{Li:2001st,Yamaguchi:2002vw, Kusenko:2014uta,Ibe:2015nfa,Takahashi:2015ula, Bae:2018mlv,Domcke:2020kcp,Berbig:2023uzs,Chao:2023ojl,Chun:2023eqc}. As long as any $L$ or effectively $B-L$ violating interaction ($e.g.,$ resulting from Weinberg operator $\ell_L \ell_L HH/\Lambda$ with $\Lambda$ as cut-off scale) stays in thermal equilibrium, the production of $B-L$ asymmetry continues until such interaction decouples from the thermal bath at temperature $T_{d}^{\rm H}$. Beyond this point, the $B-L$ asymmetry gets frozen which is later converted into the baryon asymmetry through sphaleron processes. However, in an endeavour to realise this, there are certain aspects which restrict the mechanism to take place only at very high temperature. In particular, the requirement of keeping the Weinberg operator in thermal equilibrium while the same being responsible for generating correct order of magnitude of light neutrino mass $m_{\nu}$ sets a bound $T_{d}^{\rm H} \gtrsim 10^{13}$ GeV. On the other hand, in order to attain a non-zero velocity ${\dot{\theta}}$ in the context of standard misalignment mechanism  \cite{Preskill:1982cy,Abbott:1982af,Dine:1982ah,Turner:1983he,Arias:2012az}, the ALP must start oscillating at $T_{\rm osc}$ by then, $i.e. ~T_{\rm osc} > T_{d}^{\rm H}$. Note that ALP oscillation starts when its mass ($m_\phi$) becomes comparable to the Hubble expansion rate ($\mathcal{H}$) in radiation dominated Universe, indicative of the fact that reheating temperature of the Universe after inflation $T_{\rm RH}$ also has to be larger than $T_{\rm osc}$. Such a stipulated hierarchy among the three crucial temperatures ($T_{\rm RH} > T_{\rm osc} > T_{d}^{\rm H} \sim 10^{13}$ GeV) associated to the mechanism not only demands a large value of the reheating temperature but also suggests the ALP to be heavy enough{\footnote{In case modified ALP potential \cite{Bae:2018mlv} is in place such as slow roll is incorporated, the limit becomes $m_{\phi} \gtrsim 10^5$ GeV \cite{Kusenko:2014uta,DeSimone:2016ofp}.}, $m_{\phi} \gtrsim 10^9$ GeV \cite{Ibe:2015nfa,Foster:2022ajl}.


In this work, we aim to bring down the scale of spontaneous leptogenesis attributed to much lighter ALPs (sub-GeV range) that can be probed in several ongoing and future experiments. For example, collider experiments (BaBar, CLEO, LEP, and the LHC) explore ALPs up to GeV scale via missing-energy signals \cite{CLEO:1994hzy,BaBar:2010eww}, whereas beam-dump searches and the FASER experiment in LHC are sensitive to ALPs with masses below $\mathcal{O}$(1) GeV \cite{Riordan:1987aw,Bjorken:1988as,Dobrich:2015jyk} and exceeding a few MeV \cite{FASER:2018eoc}, respectively. {Simultaneously, a realisation of such scenario with a relatively low reheating temperature would be a welcome development in view of the lower bound on $T_{RH} \gtrsim \mathcal{O}$(few MeV).}

As stated above, the high $T_{d}^{\rm H}$ is a result of a tension between the satisfaction of light neutrino mass and to keep (B-L)-violating interaction in thermal equilibrium by the same operator. We therefore propose the inclusion of another lepton number violating operator analogous 
to the Weinberg operator but replacing the SM Higgs by inert Higgs doublet (IHD) $\Phi$, $\ell_L \ell_L \Phi\Phi/\Lambda$. Being unrestricted by the neutrino mass, such IHD aided lepton number violating operator plays pivotal role in reducing the associated decoupling temperature $T_{d}^{\rm \Phi}$ while the satisfaction of neutrino mass is through the Weinberg operator. Though $T_{d}^{\rm H}$ remains unaltered (as its coefficient has to be consistent with correct neutrino mass), it is the $\ell_L \ell_L \Phi\Phi/\Lambda$ operator which can remain in thermal equilibrium till a much lower temperature, thanks to the difference in coupling coefficients in front, and allows for a lighter ALPs. Such a low temperature realisation of spontaneous leptogenesis has not been explored in the literature to the best of our knowledge. We further extend this novel platform to a next level by allowing a non-zero initial value for ALP velocity (after inflation) 
which helps bringing down the ALP mass further down to $\mathcal{O}(10)$ keV-MeV range unlike the existing literature \cite{Kusenko:2014uta,DeSimone:2016ofp}. {Although such non-zero initial ALP velocity has already been utilized within the framework of axiogenesis \cite{Co:2019wyp,Co:2020xlh,Harigaya:2021txz,Co:2022aav} and its extensions \cite{Co:2020jtv,Kawamura:2021xpu,Barnes:2022ren,Barnes:2024jap} to explain the baryon asymmetry with lighter ALPs, in this work we will focus specifically on the mechanism of spontaneous leptogenesis.}
 
 \vskip 0.1 cm

\section{Spontaneous Leptogenesis with Weinberg operator}
The presence of derivative interaction $\partial_\mu \phi j^\mu_{X}/f_\phi$ involving $j^\mu_{X} = 
{\bar{\psi}}_{{X}_{i}}\gamma^{\mu} {\psi_{X_{i}}}$ with ${\psi_{X_i}\in}$ SM lepton ($\ell_i$) and quark ($q_i$) doublets and right handed singlets of different flavors $i$) in the background of the ALP field plays a pivotal role in realising spontaneous leptogenesis. While the homogeneous nature of $\phi$ field reduces the interaction to be dependent on time derivative of $\phi$ only, the associated $j^0_X$ relates it to the number density of particles $n_{X}$ (and anti-particles ${n}_{\bar{X}}$) as 
\beeq
 \frac{c_{\scriptstyle X}}{f_\phi} (\partial_\mu \phi) j^\mu_{\scriptstyle X} \to \frac{c_X}{f_\phi}\dot{\phi} (n_{X}-{n}_{\bar{X}}), 
 \label{Egap}
\eeq
where $c_X(\in c_\ell~{\rm and}~c_q)$ is considered as a flavor-universal coupling constant (for $\ell_i$ and $q_i$, respectively).
A non-zero $\dot{\phi}$ therefore causes the above interaction to be CPT violating in nature which exhibits a shift in energy for individual particles (by $\frac{c_X}{f_\phi}\dot{\phi}$) and anti-particles (by $-\frac{c_X}{f_\phi}\dot{\phi}$) reminiscent of an effective chemical potential, $\mu_X= -\mu_{\bar{X}} =c_X{\dot{\phi}}/{f_\phi}$ where the particles and anti-particles are assumed to be in thermal equilibrium. 

The effective chemical potential $\mu_X$ thus generated acts as a seed for an $equilibrium$ number-density asymmetry 
($n_{X}^{\rm eq}-{n}_{\bar{X}}^{\rm eq}$) between leptons (quarks) and anti-leptons (anti-quarks) provided there exists a 
$L(B)$ violating interaction in thermal equilibrium. Hence it is possible to generate an equilibrium $B-L$ asymmetry 
$n^{\rm eq}_{B-L}$, expressed in terms of the chemical potential $\mu_{B-L}=(59/78)\dot{\theta}$ (estimated in \cite{Bae:2018mlv,Shi:2015zwa}) as
\beeq
n_{B-L}^{\rm eq}=(n_q^{\rm eq}-\bar{n}_q^{\rm eq} ) -(n_\ell^{\rm eq}-\bar{n}_\ell^{\rm eq}) \simeq \frac{1}{6}\mu_{B-L} T^2,
\label{nBLeq1}
\eeq
where thermal distribution of $\ell, \bar{\ell}~(q,\bar{q})$ at temperature $T < T_{\rm RH}$ is employed.

A natural choice of such $B-L$ violating interaction is the Weinberg operator \cite{Weinberg:1979sa}, responsible for neutrino mass generation, 
{\beeq
\mathcal{L}_{\sla{L}}^{\rm H}=\frac{1}{2}\kappa_{ij} \frac{(\bar{{\ell}}_{L_i} \tilde{H})(\tilde{H}^T{\ell}_{L_j}^C)}{\Lambda},
\label{weinberg}
\eeq
where $H$ is the SM Higgs doublet with the definition $\tilde{H}\equiv i \sigma_2 H^\star$, $\kappa$ is the coupling matrix, and $\ell_L$ is the SM left handed lepton doublet and $\Lambda$ is the cut-off scale. This induces neutrino mass matrix $m_{\nu} = \kappa \frac{v^2}{2\Lambda}$ after $H$ gets vev $v=246$ GeV. 


Note that this $B-L$ violating interaction remains in thermal equilibrium till a point, characterized by decoupling temperature $T_d^{\rm H}$, beyond which 
the corresponding interaction rate $\Gamma_\slashed{L}^{\rm H}$ becomes comparable (or smaller) to Hubble $\mathcal{H}  (=1.66\sqrt{g_\star}T^2/M_{\rm Pl}$ in radiation-dominated Universe, below $T_{\rm RH}$) where
$\Gamma_{\slashed{L}}^{\rm H}= 4  n_{\ell}^{\rm eq} \langle\sigma v\rangle \approx {3 n_{\ell}^{\rm eq} \sum_i m_{\nu_i}^2}/{(2\pi v^4)}$.
Here $\langle\sigma v\rangle$ denotes the thermally averaged cross section for lepton number violating processes 
$(\ell_L \ell_L\leftrightarrow HH, \ell_L H \leftrightarrow \overline{\ell}_{L},\overline{H})$ arising $solely$ from Eq.~\eqref{weinberg} and $n_\ell^{\rm eq}\approx 2T^3/\pi^2$. Using the latest fit for neutrino data \cite{deSalas:2020pgw}, 
$\sum_i m_{\nu_i}^2 \sim \Delta m^2_{atm} = 2.5 \times 10^{-3}$ eV$^2$ (considering normal hierarchy and lightest neutrino to be massless), the decoupling temperature is uniquely fixed at  
$T_{d}^{\rm H} \simeq 2\times 10^{13}$ GeV which serves as a characteristic scale that determines the final $B-L$ asymmetry 
($n_{B-L}$). This is because, below $T_{d}^{\rm H}$, $n_{B-L}^{\rm eq}$
eventually gets frozen which would further be converted into final baryon asymmetry $(n_B)$ by weak sphalerons \cite{Khlebnikov:1988sr,Arnold:1987zg,Harvey:1990qw} via
$n_B=(28/79)n_{B-L}$. A more precise estimate of the final asymmetry is to be obtained employing Boltzmann equation  \cite{Kusenko:2014uta,Bae:2018mlv}.


It is interesting to note that the other important ingredient to realise $n_{B-L}$ is related to the fact that the ALP field must have non-zero velocity ($\dot{\theta}$) before Universe reaches $T_{d}^{\rm H} \sim 10^{13}$ GeV. In the context of standard misalignment mechanism, the ALP field is assumed to be stuck at some initial value $\theta_i=\mathcal{O}(1)$ after inflation, say at reheating $T_{\rm RH}$, until the condition $3 \mathcal{H}(T_{\rm osc})= m_\phi$ is achieved after which it moves toward the minimum of its potential, hence acquiring a non-zero velocity. The  oscillation temperature $T_{\rm osc}$ followed from this relation is given by
\beeq
T_{\rm osc}\simeq 1.5\times 10^{13} \textrm{GeV}\left( \frac{100}{g_\star(T_{\rm osc})}\right)^{1/4}\left(\frac{m_{\phi}}{10^{9}\textrm{GeV}} \right)^{1/2}.
\label{Tosc1} 
\eeq
In order to fulfil all the requirements, a very restrictive range of high temperature emerges, $T_{\rm RH} > T_{\rm osc} > T_{d}^{\rm H} 
\sim 10^{13}$ GeV, to realise spontaneous leptogenesis. Furthermore, as shown in \cite{Shi:2015zwa}, unless $T_d^{\rm H}$ can be lowered below the temperature where the weak sphaleron enters  equilibrium ($T_{\rm ws} \sim 10^{12}$ GeV) \cite{Bento:2003jv}, the residual asymmetry would vanish.

\vskip 0.2cm
\section{Spontaneous Leptogenesis with Inert Higgs Doublet}
The relation of Eq.~\eqref{Tosc1} is certainly indicative of heavy ALPs ($m_\phi \gtrsim10^9$ GeV) while, as stated in 
the introduction, a light ALP would be interesting from experimental viewpoint. Also, the reheating temperature can in principle be substantially lower than $10^{13}$ GeV~\cite{Giudice:2000ex,Kawasaki:2000en,Martin:2010kz,Dai:2014jja,Datta:2022jic} (depending on the coupling of inflaton with SM fields)
and in that case, the above scenario would no longer in use.  
We note that the main obstacle to realise the spontaneous leptogenesis at a lower temperature follows from the necessity of $B-L$ breaking interaction originated from Weinberg operator to 
remain in thermal equilibrium via $\Gamma_{\slashed{L}}^{\rm H} \gtrsim \mathcal{H}$ condition, which is also intricately 
tied up with neutrino mass generation. As a resolution to this problem, we propose to include another $B-L$ violating operator, 
{\beeq
\mathcal{L}_{\sla{L}}^{\rm \phi}=\frac{1}{2} \frac{(\bar{{\ell}}_{L_i} \tilde{\Phi})(\tilde{\Phi}^T{\ell}_{L_j}^C)}{\Lambda}
,~\text{with}~~ \Phi =
\begin{bmatrix}
	\Phi^{+}\\
	\Phi^0
\end{bmatrix} ~\&~\tilde{\Phi}\equiv i \sigma_2 \Phi^\star
\label{weinberg-IHD}
\eeq}
analogous to Weinberg operator, replacing the SM Higgs by an IHD $\Phi$ which does not carry any $vev$. 
The IHD being secluded from neutrino mass generation\footnote{It can be noted that a contribution to light neutrino mass may also originate at two-loop using both the dimension-5 operators (Eqs. (\ref{weinberg}) and (\ref{weinberg-IHD})). However, due to the involvement of an effective coefficient $\sim 1/\Lambda^3$, its contribution will be highly suppressed.} has the potential to allow a greater flexibility between the decoupling temperature of the relevant $B-L$ violating interaction ($T_d^\Phi$ associated to $\mathcal{L}_{\sla{L}}^{\Phi}$) and 
reheating temperature $(T_{\rm RH})$. Inclusion of IHD brings an additive benefit in terms of its contribution to dark matter candidate (protected by the $Z_2$ symmetry). While the involvement of gauge coupling in case of IHD ensures its presence in the thermal bath at an early Universe, it is well known that the lightest neutral component of the IHD can play the role of a freeze-out type of dark matter. Several studies show that in principle there would be two mass regimes of the IHD as DM, one is below 80 GeV
and the other is above 550 GeV, for which the relic density and direct detection limits are satisfied \cite{Barbieri:2006dq,Ma:2006km,LopezHonorez:2006gr,LopezHonorez:2010eeh,Borah:2012pu,Gustafsson:2012aj,Biswas:2013nn,Arhrib:2013ela,Keus:2014jha,Ilnicka:2015jba,Belyaev:2016lok,Choubey:2017hsq,Kalinowski:2018ylg,Bhardwaj:2019mts,Bhattacharya:2019fgs,Borah:2019aeq,DuttaBanik:2020vfr}. 
Such mass regimes are not affected by the additional dimension-5 interaction of Eq. (\ref{weinberg-IHD}) we considered, primarily because of the suppressed nature of the interaction.

In presence of both the $B-L$ violating operators, we first observe that $T_d^{\rm H}$ remains essentially unchanged as the associated interaction rate $\Gamma_{\slashed{L}}^{\rm H}$ solely depends on neutrino mass ($i.e.$ independent to $\Lambda$). On the other hand, interaction rate associated to $\mathcal{L}_{\sla{L}}^{\rm \Phi}$ being ${\Gamma_{\slashed{L}}^{\rm \Phi}=
{3g n_{\ell}^{\rm eq}}/{(8\pi\Lambda^2)}}$ with $g=324/23$ (see Appendix~\ref{section2}), 
$T_d^{\rm \Phi}$ can be made much smaller than $T_d^{\rm H}$ as
\beeq
T_d^{\rm \Phi}\simeq 4 \times 10^{6} ~{\rm GeV} \left(\frac{g_\star}{100} \right)^{1/2} \left(\frac{\Lambda}{10^{12}~{\rm GeV}} \right)^{2},
\label{Td-Lambda}
\eeq
signifying that the IHD assisted interaction may persist in thermal equilibrium 
for a prolonged period than the case with Weinberg operator ($\mathcal{L}_{\sla{L}}^{\rm H}$) alone. 
As shown in Fig. \ref{Tdec-lam}, the decoupling temperature associated to the new interaction, $T_d^{\Phi}$, can 
be reduced 
by following the relation of Eq. (\ref{Td-Lambda}). Note that with any such $\Lambda$ corresponding to a particular $T_d^{\Phi}$, the neutrino mass can be made of right order by adjusting the coupling parameter $\kappa$ involved in $m_\nu=\kappa\frac{v^2}{2\Lambda}$. 
However, in order the elements of this coupling matrix should satisfy $\kappa_{ij} \lesssim 1$, a restriction on the higher value of $\Lambda$ follows as seen in Fig. \ref{Tdec-lam}.

\begin{figure}[]
	\hspace{-0.2cm}
	\includegraphics[scale=0.6]{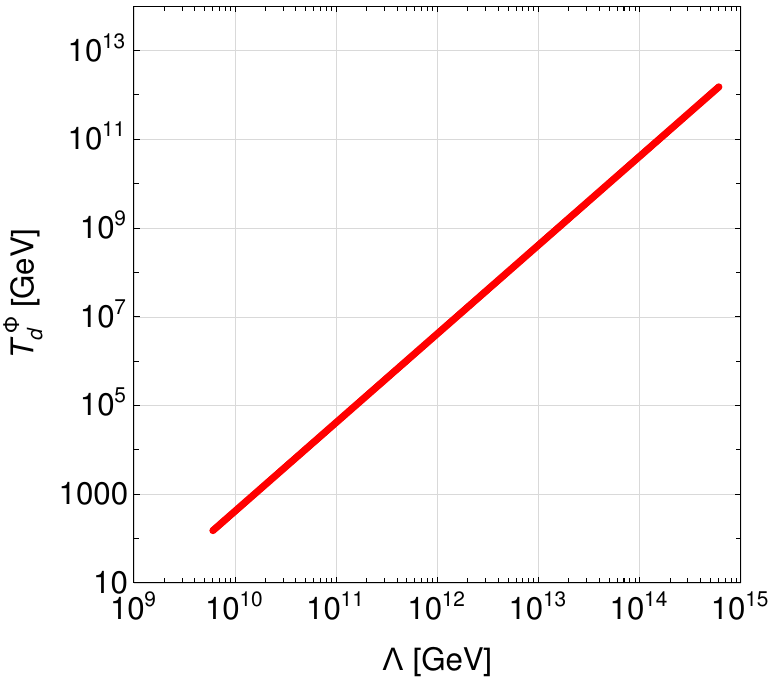}
	\caption{Variation of decoupling temperature $T_d^{\rm \Phi}$ (GeV) against the cut-off scale $\Lambda$ (GeV).}
	\label{Tdec-lam}
\end{figure}

Before estimating the $B-L$ asymmetry in this case, let us analyse the ALP dynamics starting from $T_{\rm RH}$ in order to estimate its velocity $\dot{\theta}$ which is crucial in determining $n_{B-L}$. The global $U(1)$ symmetry is considered to be broken before inflation rendering the ALP as a homogeneous field having effective potential $V(\phi)=m_\phi^2 f_\phi^2 \left(1-\textrm{cos}\frac{\phi}{f_\phi}\right)$ which obeys the following equation of motion, 
\beeq
\ddot{\phi}+ 3 \mathcal{H} \dot{\phi}+\frac{\partial V(\phi)}{\partial \phi}= \frac{c_X}{f_\phi a^3} \partial_t (a^3 j_X^0).
\label{ALP-eom1}
\eeq
The r.h.s of Eq.~\eqref{ALP-eom1} representing a back-reaction term, originated from $ c_X(\partial_\mu \phi) j^\mu_{X}/f_\phi \label{leff-B-L}$, can be safely excluded while studying the ALP evolution for $T < T_{\rm RH}$ with the consideration $f_\phi\gtrsim T_{\rm RH}$. 
However, solving Eq.~\eqref{ALP-eom1} requires to specify initial conditions associated to $\phi$. We set the initial field value of ALP, $\phi_i$, to be $\phi_i=\phi(T_{\rm RH})=f_\phi$ (equivalently, $\theta_i= 1$). In addition, initial condition on $\dot{\theta}_i$ can be set to zero (referred as case A) under conventional misalignment mechanism \cite{Preskill:1982cy,Abbott:1982af,Dine:1982ah,Turner:1983he,Arias:2012az} or non-zero (case B) within the so-called kinetic misalignment mechanism \cite{Co:2019wyp,Co:2019jts,Chang:2019tvx,Co:2020jtv,Chakraborty:2021fkp,Kawamura:2021xpu,Barman:2021rdr,Co:2022aav,Barnes:2022ren,Barnes:2024jap} as discussed below. 

\vskip 0.3 cm 

\noindent [{\bf{A: Freeze-in scenario}}] Primarily, with $\theta_i=\mathcal{O}(1)$ and $\dot{\theta}_i$ = 0, ALP would start oscillating at $T_{\rm osc}$ defined by Eq.~\eqref{Tosc1} and gain non-zero $\dot{\theta}$ obtainable using the solution of Eq.~\eqref{ALP-eom1}. With the approximated ALP potential near minimum $V(\theta)\simeq m_\phi^2 f_\phi^2 \theta^2/2$, solution of Eq. (\ref{ALP-eom1}) can approximately (neglecting r.h.s.) be given by 
\beeq
\theta(t)\simeq \theta_i~ \Gamma\left(\frac{5}{4}\right)\left(\frac{2}{m_\phi t}\right)^{1/4} J_{1/4}(m_\phi t),
\eeq
in radiation-dominated Universe, where $J_{1/4}$ refers to the Bessel's function of first kind. From $\theta(t)$, the ALP velocity 
$\dot{\theta}$ can easily be estimated at any point of time. Provided $T_{\rm osc} > T_{d}^{\rm \Phi}$ can be realised with suitable choices of $m_{\phi}$ and $\Lambda$ (as in Table \ref{BP-table}), $n_{B-L}^{\rm eq}$ is obtained using Eq. (\ref{nBLeq1}) with $\mu_{B-L} = -(8/3)\dot{\theta}$ in our case (see Appendix~\ref{section1}).
Then a precise estimate of final $n_{B-L}$ results by solving the Boltzmann equation
\beeq
\dot{n}_{B-L}+3 \mathcal{H} n_{B-L}=-\Gamma_{\slashed{L}}^{\rm \Phi} \left(n_{B-L}-n_{B-L}^{\rm eq}\right). 
\label{beqn-IHD}
\eeq 
through decoupling epoch.



{Fig. \ref{BP-caseA} (upper panel) demonstrates the evolution of the equilibrium $B-L$ asymmetry $Y_{B-L}^{\rm eq}=n_{B-L}^{\rm eq}/s$ (indicated by the red curve) and the resultant $B-L$ asymmetry $Y_{B-L}=n_{B-L}/s$ (marked in blue line)  obtained as a solution to Eq. (\ref{beqn-IHD}), against normalised scale factor $a/a_{\rm end}$ ($a_{\rm end}$ being the scale factor at the end of inflation, set at $a/a_{\rm end}=1$) for benchmark point BP1, mentioned in Table \ref{BP-table}. 
As can be seen from Fig. \ref{BP-caseA}, the $Y_{B-L}^{\rm eq}$, being proportional to the ALP velocity $\dot{\theta}$, tracks the evolution of $\theta$. Hence, $Y_{B-L}^{\rm eq}$, initially starting from zero in this case, oscillates following the ALP oscillation at a temperature $T \leq T_{\rm osc}$. On the other hand, while the $B-L$ violating interaction prevails in thermal equilibrium, $Y_{B-L}$ traces closely the $Y_{B-L}^{\rm eq}$ and starts to grow} 
gradually from zero once the ALP field starts moving toward its minimum marking the onset of oscillation and reaches a peak value when the ALP field initially crosses $\theta=0$, attaining maximum velocity. Subsequently, the ALP oscillation amplitude gets red-shifted by $(T/T_{\rm osc})^{3/2}$ and the asymmetry {\it freezes in} (as $B-L$ violating operator decouples) at correct $Y_B\approx 8.7\times 10^{-11}$ value~\cite{Aghanim:2018eyx}.
We find that ALPs with $m_{\phi} \gtrsim 5\times 10^{4}$ GeV can accurately reproduce the baryon asymmetry in this case. For further lighter $m_{\phi}$, sufficient amount of $B-L$ asymmetry would not result as the ALP velocity $\dot{\theta}$, related to its mass, can't be made arbitrarily large. Below we provide an alternate scenario where a low-mass ALP (in sub-GeV regime) can successfully accompanied by correct baryon asymmetry. 
\vskip 0.1 cm

\noindent [{\bf{B: Freeze-out scenario}}] Contrary to case-A, here we propose to attribute a significantly large initial velocity to ALPs $\dot{\theta}_i \neq 0$, {the origin of which can be connected to an explicit breaking of the $U(1)$ symmetry, 
$e.g.$ considered in kinetic misalignment mechanism \cite{Co:2019jts,Chang:2019tvx}.} 
Such a velocity is however bounded by 
\beeq
|\dot{\theta}_i|\lesssim\mathcal{O}(1)\frac{ T_{\rm RH}^2}{f_\phi},
\label{thdot}
\eeq
condition, which follows from the argument that at $T= T_{\rm RH}$, ALP's kinetic energy ($\dot{\theta}^2f_\phi^2/2$) remains sub-dominant compared to the energy density of the Universe (${\pi^2}g_\star  T_{\rm RH}^4/30$). 
\begin{center}
	\begin{table}[h]
		\begin{tabular}{ |c|c|c|c|c| } 
			\hline
			BP &	$\Lambda$ (GeV)& $T_{\rm RH}$ (GeV) & $m_\phi$ (GeV) & $\dot{\theta}_i$     \\
			\hline
			\hline
			[A] BP1 &	$ 1.02\times10^{14}$& $7.4\times10^{11}$ & $7\times 10^{4}$ & $0$   \\ 
						\hline	
			[B] BP2 & $5.25\times10^{12}$& $4.5\times10^{9}$ & $1$ & $-10^5m_\phi$   \\
			\hline	
		\end{tabular} 
		\caption{Benchmark Points (BPs) for case A (standard misalignment) and case B (kinetic misalignment).}
		\label{BP-table}
	\end{table}
\end{center} 
Also, if the initial kinetic energy of ALP is larger than the height of the ALP potential ($2 m_\phi^2 f_\phi^2$), the ALP field will leap across the potential minima till a point where they become equal and ALP field being trapped in a specific minimum starts performing the oscillation. Thus actual ALP oscillation commences at $T^\star_{\rm osc}$, governed by 
$\dot{\theta}(T^\star_{\rm osc})=2m_\phi$.
However, for kinetic misalignment, $T^\star_{\rm osc}>T_d^{\rm \Phi}$ is no more a necessary condition as ALP starts evolving with an existing chemical potential at $T=T_{\rm RH}$ itself, given by non-zero $\dot{\theta}_i$. Hence, to generate the desired baryon asymmetry using $\mathcal{L}_{\sla{L}}^{\rm \Phi}$, the necessary conditions $  T_{\rm RH}>T_d^{\rm \Phi}$ and $T_{\rm RH}>T^\star_{\rm osc}$ should be met. 

\begin{figure}[!h]
	\includegraphics[scale=0.6]{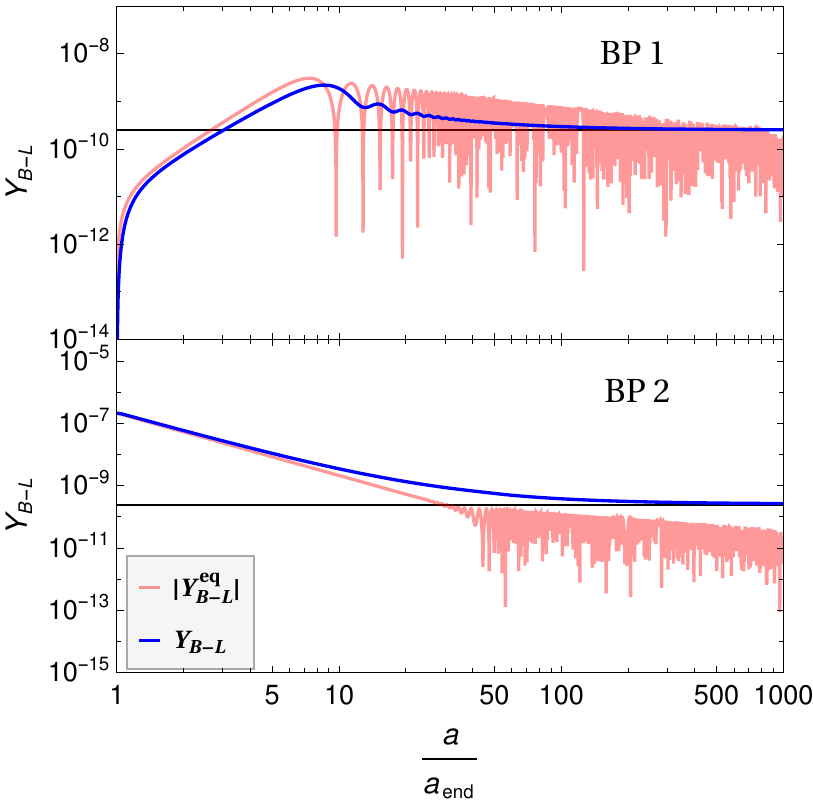}
	\caption{{\it Freeze in} ({\bf upper panel}) and {\it Freeze out} ({\bf lower panel}) of $B-L$ asymmetry  $Y_{B-L} (|Y_{B-L}^{\rm eq}|)$ displayed against $a/a_{\rm end}$ for BP1 and BP2, respectively, with solid blue line (oscillating light red shade). The black gridline represents correct value of $Y_{B-L}$ which generates $Y_B\approx 8.7\times 10^{-11}$.}
	\label{BP-caseA}
\end{figure}

Here we observe from Fig. \ref{BP-caseA} (bottom panel) that highest value of asymmetry emerges right at the beginning due to large initial $\dot{\theta}_i$ (see Table \ref{BP-table}, BP2) as $Y_{B-L}$ follows $Y_{B-L}^{\rm eq}$. Note that $T_{\rm RH}$ being smaller than $T_d^{\rm H}$, the $\ell_L \ell_L HH$ interaction does not contribute to it. 
The $B-L$ asymmetry finally {\it freezes out} as the $\ell_L \ell_L \Phi\Phi$ interaction decouples from equilibrium. We perform a parameter space scan involving $m_\phi, T_{\rm RH}$ and $\Lambda$ having initial conditions: $\theta_i=1$ and $\dot{\theta}_i=-10^5 m_\phi$ satisfying correct baryon asymmetry. The result is depicted in Fig. \ref{ps-caseB} in $m_\phi-T_{\rm RH}$ plane (with $\Lambda$ in the colour-bar\footnote{With each $\Lambda$, there exists $\kappa$ followed from $m_{\nu} = \kappa v^2/(2\Lambda)$.}) focusing on an extended range of ALP mass in the sub-GeV realm associated to $T_{\rm RH} < {\mathcal{O}}(10^9)$ GeV. While the lower limit of the scan remains robust with this initial condition, achieving correct  BAU with a much lighter ALP remains plausible by using a further larger $\dot{\theta}_i$.
\begin{figure}[t]
	\centering
	\includegraphics[scale=0.48]{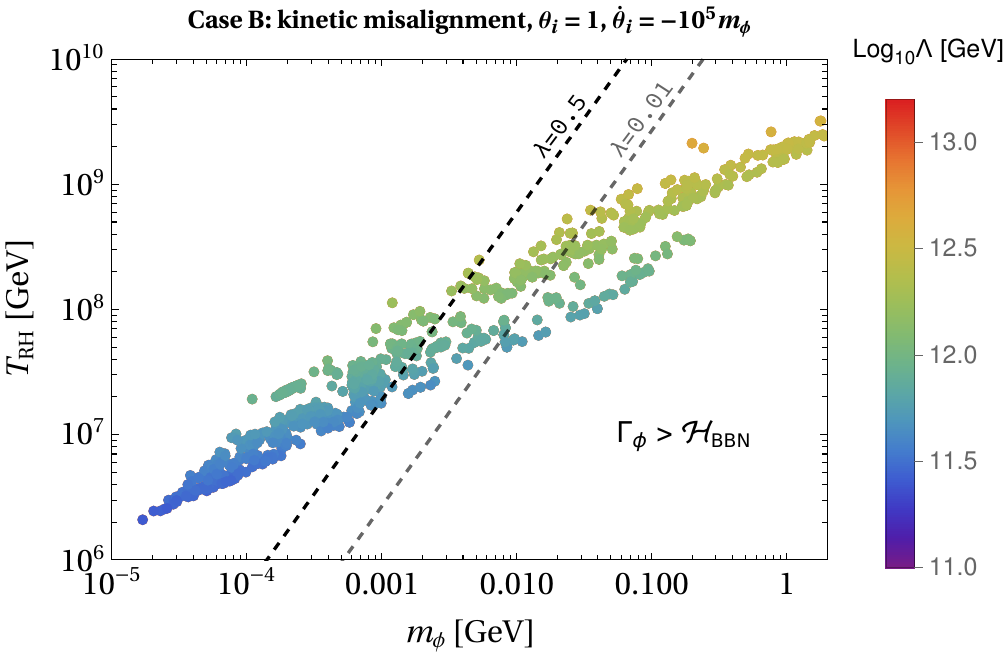}
	\caption{Parameter scan for case B: $\dot{\theta}_i\neq 0$ in $m_\phi-T_{\rm RH}$ plane. Here $\dot{\theta}_i=-10^5 m_\phi$ is considered. In the color map, the parameter $\Lambda$ is specified in log-scale.}
	\label{ps-caseB}
\end{figure}

Finally, we address the fate of these ALPs at a later stage, following the freeze-in or freeze-out of the $B-L$ asymmetry, as it continues to oscillate thereafter. 
{It is pertinent to note that even if ALP possesses a large initial velocity (as in case B), the kinetic energy of ALPs being scaled as $\rho_\phi^{\rm K.E.}\propto a^{-6}$ will quickly die out and become negligibly small in comparison to its potential energy density $(\propto a^{-3})$ as well as $\rho_R~(\propto a^{-4})$. This results into trapping of the ALP within one potential well within which it starts to oscillate. Therefore its energy density is thereafter dominated by its potential energy which scales as $\rho_\phi (T) \propto a(T)^{-3}$. As a result of which, although the energy density of ALP are set to be sub-dominant compared to the thermal bath at the time of reheating, a possibility may open up where the ALP can dominate the energy density of the Universe at the time of BBN.
Hence, it is preferable that they should decay (with decay rate $\Gamma_{\phi}$) prior to BBN so as not to disturb BBN prediction, $i.e.~\Gamma_{\phi} \gtrsim \mathcal{H}_{\rm BBN}$ \cite{DeSimone:2016ofp,Kawasaki:2000en,Hasegawa:2019jsa}}. ALPs are primarily expected to decay into SM leptons as well as $WW/ZZ$ (provided kinematically allowed) via 
their effective interactions \cite{Georgi:1986df} which can be parametrised by $\Gamma_{\phi}=\beta{m_\phi^3}/{f_\phi^2}$, with $\beta < \mathcal{O}$(1). Using this along with the criteria considered $T_{\rm RH} = \alpha  f_\phi$ where $\alpha \leq 1$, the residual allowed parameter space falls in right side of black dashed $T_{\rm RH} - m_{\phi}$ contour lines (acting as boundary of $\Gamma_{\phi} \geq \mathcal{H}_{\rm BBN}$) for specific choices of $\lambda 
~(= \beta \alpha^2) <1$ in Fig. \ref{ps-caseB} that could be sensitive to upcoming ALPs searches. The constraints on such decaying ALP in the mass regime of Fig. \ref{ps-caseB} also stem from astrophysics (e.g. {stellar evolution \cite{Carenza:2020zil} as well as emission and decays of ALP out of supernova SN1987A \cite{Jaeckel:2017tud,Lucente:2020whw,Caputo:2022mah,Hoof:2022xbe}}), electron and proton beam dump experiments ({e.g. SLAC E137 \cite{Bjorken:1988as,Dolan:2017osp},  CHARM and NuCal \cite{Dobrich:2015jyk} etc.}) and collider experiments {like CLEO and BaBar \cite{CLEO:1994hzy,BaBar:2010eww}, LEP-I and II \cite{L3:1994shn,OPAL:2002vhf,Knapen:2016moh}, Belle-II \cite{Belle-II:2020jti}, PrimEx \cite{Aloni:2019ruo} etc.}, which are shown in Fig. \ref{ps-caseB-agg} considering the ALP-photon interaction given by ${g_{\phi\gamma\gamma}}\phi F\tilde{F}/{4}$. The ALP-photon coupling $g_{\phi\gamma\gamma}$ is related to the ALP decay constant $(f_\phi)$ as
\beeq
g_{\phi\gamma\gamma}=\frac{\alpha}{2\pi f_\phi}C_{\phi\gamma\gamma},
\eeq 
where $C_{\phi\gamma\gamma}$ is considered to be $\mathcal{O}(1)$ and $\alpha$ refers to the fine structure constant. Note that $f_\phi$ is not directly involved in the evolution of baryon asymmetry except the requirement $f_\phi\gtrsim T_{\rm RH}$ (see discussion related to Eq. \eqref{ALP-eom1}). Keeping this in mind, we choose two specific relations among $f_\phi$ and $T_{\rm RH}$ to demonstrate the parameter space obtained in Fig. \ref{ps-caseB} onto $m_\phi-g_{\phi\gamma\gamma}$ plane (in Fig. \ref{ps-caseB-agg}) as (a) $f_\phi=T_{\rm RH}$ (marked with triangular points) and (b) $f_\phi=100~ T_{\rm RH}$ (marked with cross points). As seen from Fig. \ref{ps-caseB-agg}, for both choices of $f_\phi$, the allowed parameter space in our case lies below the current exclusion limits from both astrophysics and experiments (collider and beam dump), which can be probed in future.

\begin{figure}[t]
	\centering
	\includegraphics[scale=0.48]{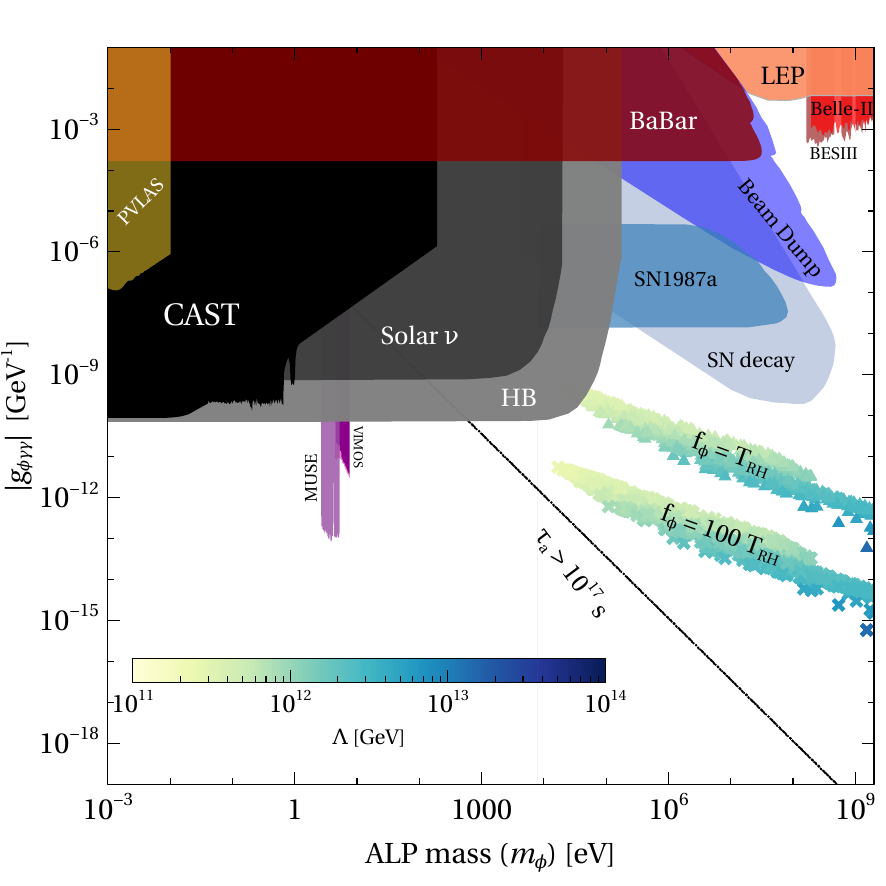}
	\caption{ALP parameter space in $m_\phi-g_{\phi\gamma\gamma}$ plane for case B ($\dot{\theta}_i=-10^5 m_\phi$) with exclusion regions from various constraints. In the color map, the parameter $\Lambda$ is specified in log-scale.}
	\label{ps-caseB-agg}
\end{figure}

\vskip 0.1 cm

\section{Summary}
In summary, we propose a unique scenario of leptogenesis for generating the BAU by incorporating a dynamic CPT violating effect involving light ALPs in presence of a thermally equilibrated lepton number violating interactions in the early Universe. Traditionally, such possibility comprises lepton number violating dimension-5 Weinberg operator that simultaneously can account for neutrino mass. However, this being in thermal equilibrium only at very high temperature $(T \gtrsim10^{13} ~{\rm GeV})$ in the early Universe, such mechanism turns out to be redundant both for the low reheating scenarios and those with light ALPs, which are otherwise interesting from experimental point of view. This work resolves both these downsides at one go, hitherto unexplored in the literature, by introducing a $B-L$ breaking operator (analogous to the Weinberg operator) involving IHD instead of the SM Higgs. Using the freedom associated with this new operator, $i.e.$ not being connected to neutrino mass, the decoupling temperature of lepton number violating interactions can be significantly lowered. This enables the generation of correct BAU via $\it{freeze}$-$\it{in}$ with low $T_{\rm RH}$ while simultaneously accounts for lighter ALPs $m_{\phi} \sim 5\times 10^{4}$ GeV even without any initial ALP velocity. The study further extends to lower the mass ALPs down to $\sim$ $\mathcal{O}(10)$ keV-MeV leading to $\it{freeze}$-$\it{out}$ production of BAU, where a large initial velocity of ALP is considered making it intriguing for search of ALPs. Moreover, the IHD here can serve as a potential potential dark matter candidate, bridging the connection with another unresolved problem securing minimality of the construction.

\begin{acknowledgements}
AD acknowledges the National Research Foundation of Korea (NRF) grant funded by the Korean government (2022R1A5A1030700) and the support provided by the Department of Physics, Kyungpook National University. AD would also like to thank Kyu Jung Bae for useful discussions on axion dynamics. The work of AS is supported by the grants CRG/2021/005080 and MTR/2021/000774 from SERB, Govt. of India. 
\end{acknowledgements}

\onecolumngrid

\appendix

\section{Relationship among chemical potentials in presence of IHD}
\label{section1}
Here, we proceed to evaluate first the effective chemical potential $\mu_{B-L}$ contributing toward the equilibrium number density of $B-L$ charges 
$n_{B-L}^{\rm eq}$ in presence of the inert Higgs doublet (IHD) in section~\ref{section1}. Although the basic prescription can be found in \cite{Shi:2015zwa, Bae:2018mlv}, 
the inclusion of IHD would alter the final result.

During the evolution of the Universe, fundamental particles participating in an interaction (say, $i+j+...\leftrightarrow a+b+..$) remains in chemical equilibrium as long as the respective interaction rate becomes comparable to the Hubble expansion rate of the Universe. When in equilibrium, the associated chemical potentials of the participating particles would follow a relation
\begin{align}
	\mu_i+\mu_j+...+\mu_a+\mu_b+...=0.
\end{align}
As a consequence, one can have inter-relations among chemical potentials of different particles leading to an estimate for $\mu_{B-L}$ as discussed below. 
\begin{itemize}
	
	\item The gauge interactions come into equilibrium at very early Universe where the electroweak symmetry remain unbroken. In this unbroken phase, the gauge bosons remain massless leading to vanishing chemical potentials, $\mu_B=\mu_W=\mu_g=0$. As a consequence, the components of the electroweak and color multiplets will have the same chemical potentials i.e. $\mu_q\equiv \mu_{u_L}=\mu_{d_L}$, $\mu_l\equiv \mu_{\nu_L}=\mu_{e_L}$, and $\mu_H\equiv\mu_{H^+}=\mu_{H^0}$. Similarly, the chemical potential of the additional IHD also follows $\mu_\Phi\equiv\mu_{\Phi^+}=\mu_{\Phi^0}$.
	
	\item IHD being a scalar, has self interactions (consistent with its $Z_2$ odd property for being a dark matter candidate) as well as portal interactions with the Standard Model (SM) Higgs doublet~\cite{Barbieri:2006dq,Ma:2006km,LopezHonorez:2006gr}. One such term in these portal interactions is $\lambda(\Phi^\dagger H)^2$ whose chemical equilibration leads to 
	$\mu_\Phi=\mu_H$. As a consequence, the hypercharge neutrality condition takes the form
	\begin{align}
		\mu_q+2\mu_u -\mu_d-\mu_l -\mu_E+\frac{4}{N_f}\mu_H=0,
		\label{eq2}
	\end{align}
	where $\mu_{u,d (E)}$ are the chemical potential of the SM singlet right handed quarks (leptons). Here $N_f$ (=3 in SM) represents the number of generations of the quarks and leptons.
	\item Since we include a low reheating temperature ($T_{\rm RH} \ll 10^{12}$ GeV) in the study, all $N_f$ generations of Yukawa interactions are considered to be in equilibrium\footnote{For an extended reheating phase, see \cite{Datta:2022jic,Datta:2023pav} for details.}. This leads to the following relations
	\begin{align}
		\mu_u=\mu_q+\mu_H,\qquad \mu_d=\mu_q-\mu_H,\qquad \mu_E=\mu_l-\mu_H.
		\label{eq3}
	\end{align}
	\item At $T \lesssim 10^{12}$ GeV, electroweak (EW) sphalerons enter in thermal equilibrium~\cite{Bento:2003jv}, enforcing $3\mu_q+\mu_l= 0$. This is going to be modified in presence of CPT violating interactions as discussed below. 
\end{itemize}

\noindent In our framework, the CPT violating interaction of the form $\frac{c_X}{f}\partial_\mu \phi J^\mu_X$ (in a time dependent and homogeneous ALP background)  induces \cite{Shi:2015zwa} an additional effective chemical potential $\bar{\mu}_{\rm ws}$ to the chemical potential relation describing the EW-sphaleron
process, originated from chiral anomaly, given by
\begin{align}
	3\mu_q+\mu_l= -\bar{\mu}_{\rm ws}.
	\label{eq4}
\end{align}
Furthermore, the same CPT violation also modifies the chemical potential relation associated to the $B-L$ violating operator $\ell_L\ell_L \Phi\Phi/\Lambda$ (when in equilibrium) as \cite{Shi:2015zwa}:
\begin{align}
	\mu_l+\mu_\phi=\mu_l+\mu_H= -\bar{\mu}_{\ell}.
	\label{eq5}
\end{align}

Given all the SM leptons and baryons are in thermal and chemical equilibria, $n_{B-L}^{\rm eq}$ can then be calculated from $n_{B-L}^{\rm eq}=\mu_{B-L}T^2/6$ , where $\mu_{B-L}$ takes the form 
\begin{align}
	\mu_{B-L}=
	\mu_B-\mu_L
	=(2\mu_q+\mu_u+\mu_d)N_f-(2\mu_l+\mu_E)N_f\\
	=N_f\left(-\frac{4}{3}\bar{\mu}_{\rm ws}+\frac{13}{3}\bar{\mu}_{\ell}+\frac{16}{3}\mu_H\right),
\end{align}
obtained by employing Eqs. (\ref{eq3})-(\ref{eq5}).
Finally, rewriting $\mu_H$ in terms of $\bar{\mu}_{\rm ws}$ and $\bar{\mu}_{\ell}$ through Eq. (\ref{eq2}), we obtain the equilibrium $\mu_{B-L}$ in terms of $\bar{\mu}_{\rm ws}$ and 
$\bar{\mu}_{\ell}$ as 
\begin{align}
	\mu_{B-L}=-\frac{4N_f(1+N_f)}{3+5N_f}\bar{\mu}_{\rm ws}+\frac{N_f(11N_f-13)}{(3+5N_f)}\bar{\mu}_{\ell}.
\end{align}
However, as shown in \cite{Shi:2015zwa}, with proper rotations on the quarks and leptons, we can transform the basis to $\left\lbrace \bar{\mu}_{\rm ws},\bar{\mu}_{\ell} \right\rbrace \to  \left\lbrace \delta_{\rm ws},0\right\rbrace$ with $\delta_{\rm ws} =  \dot{\theta}$, the non-zero ALP velocity \cite{Bae:2018mlv}. In the new basis, $\mu_{B-L}$ can be re-written as
\beeq
\mu_{B-L}=\frac{4N_f(1+N_f)}{3+5N_f}\delta_{\rm ws}.
\label{eq9}
\eeq

\section{Construction of the Boltzmann equation}\label{section2}
The modified Boltzmann equation relevant for the $B-L$ asymmetry evolution is constructed here.
To study the evolution of the produced $B-L$ asymmetry across the decoupling of $B-L$ violating interaction (associated to  $\ell_L\ell_L\Phi \Phi/\Lambda$), a Boltzmann equation 
for $n_{B-L}$ can be constructed having the following form, 
\beeq
\dot{n}_{B-L}+3 H n_{B-L}= \frac{\gamma_{\slashed{L}}}{T}\sum_{i,j=3}(\mu_{l_i}+\mu_{l_j}+2 \mu_H)=18 \frac{\gamma_{\slashed{L}}}{T} (\mu_l+\mu_H),
\eeq 
assuming flavour-universal interaction rate density $\gamma_{\slashed{L}}$ for different leptons. In the r.h.s of the Boltzmann equation, $\mu_l$ and $\mu_H$ can be expressed 
in terms of $n_B$ and $n_L$ as
\begin{align}
	\mu_l&=\frac{24(1+N_f)n_L - 3 N_f n_B  }{2N_f(6+5 N_f) T^2}, \label{eq11}\\
	\mu_H&=\frac{12 n_L-9 n_B}{2(6+5 N_f)T^2}.\label{eq12}
\end{align}
The $n_B$ and $n_L$ can be further related to the chemical potential of the weak sphaleron $\delta_{ws}$ and $n_{B-L}$ by employing Eqs. (\ref{eq2})-(\ref{eq5}) and can be denoted as
\begin{align}
	n_B&=\frac{T^2}{6} 4 N_f \mu_q=\frac{12(1+N_f) n_{B-L}-\delta_{\rm ws}T^2 N_f(6+5 N_f)}{39+33 N_f};\label{eq13}\\
	n_L&= \frac{T^2}{6} N_f(2 \mu_l+\mu_E)= -\frac{3 (9+7 N_f)n_{B-L}+\delta_{\rm ws}T^2 N_f(6+5 N_f)}{39+33 N_f}.\label{eq14}
\end{align}
Finally, substituting Eqs. (\ref{eq13}) and (\ref{eq14}) in both the Eqs. (\ref{eq11}) and (\ref{eq12}), we obtain the final form of the Boltzmann equation as
\begin{align}
	\dot{n}_{B-L}+3 H n_{B-L}&= 
	-18\frac{6(3+5 N_f)}{N_f(13+11 N_f)}\frac{\gamma_{\slashed{L}}}{T^3}\left[n_{B-L}+\frac{4N_f(1+N_f)}{6(3+5 N_f)}\delta_{\rm ws}T^2\right],\\
	&=-\Gamma_{\slashed{L}}^{\Phi}(n_{B-L}-n_{B-L}^{\rm eq}),
\end{align}
where $\Gamma_{\slashed{L}}^\Phi$ is given by,
\beeq
\Gamma_{\slashed{L}}^\Phi=18\frac{6(3+5 N_f)}{N_f(13+11 N_f)}\frac{\gamma_{\slashed{L}}}{T^3}.
\label{mbl,glp}
\eeq
Note that with $N_f=3$, Eqs.~(\ref{eq9}) and (\ref{mbl,glp}) become $\mu_{B-L}= -\frac{8}{3} \dot{\theta}$ and $\Gamma_{\slashed{L}}^\Phi= \frac{324}{23} \frac{\gamma_{\slashed{L}}}{T^3} \simeq  \frac{324}{23} \frac{6 T^3}{8\pi^3 \Lambda^2}$ respectively, which we used in our analysis.  
\twocolumngrid
\bibliography{ref.bib}

\end{document}